\documentclass[sigconf]{acmart}

\usepackage{booktabs}
\usepackage{multirow}

\copyrightyear{2017}
\acmYear{2017}
\setcopyright{acmcopyright}
\acmConference{WebSci '17}{June 25-28, 2017}{Troy, NY, USA}
\acmPrice{15.00}
\acmDOI{http://dx.doi.org/10.1145/3091478.3091504}
\acmISBN{978-1-4503-4896-6/17/06}

\begin{document}

\title{Sharing Means Renting?: An Entire-marketplace Analysis of Airbnb}

\author{Qing Ke}
\affiliation{%
\institution{Indiana University, Bloomington}
}
\email{qke@indiana.edu}

\begin{abstract}
Airbnb, an online marketplace for accommodations, has experienced a staggering
growth accompanied by intense debates and scattered regulations around the
world. Current discourses, however, are largely focused on opinions rather than
empirical evidences. Here, we aim to bridge this gap by presenting the first
large-scale measurement study on Airbnb, using a crawled data set containing
$2.3$ million listings, $1.3$ million hosts, and $19.3$ million reviews. We
measure several key characteristics at the heart of the ongoing debate and the
sharing economy. Among others, we find that Airbnb has reached a global yet
heterogeneous coverage. The majority of its listings across many countries
are entire homes, suggesting that Airbnb is actually more like a rental
marketplace rather than a spare-room sharing platform. Analysis on star-ratings
reveals that there is a bias toward positive ratings, amplified by a bias
toward using positive words in reviews. The extent of such bias is greater than
Yelp reviews, which were already shown to exhibit a positive bias. We
investigate a key issue---commercial hosts who own multiple listings on
Airbnb---repeatedly discussed in the current debate. We find that their
existence is prevalent, they are early-movers towards joining Airbnb, and their
listings are disproportionately entire homes and located in the US. Our work
advances the current understanding of how Airbnb is being used and may serve as
an independent and empirical reference to inform the debate.
\end{abstract}

\ccsdesc[500]{Information systems~Reputation systems}
\ccsdesc[500]{Applied computing~Law, social and behavioral sciences}

\keywords{Airbnb; measurement; sharing economy; online marketplace}

\maketitle

\section{Introduction} \label{sec:intro}

Recent years have seen a proliferation of online peer-to-peer
marketplaces~\cite{Einav-p2p-2016}. Examples abound and the services covered
range from personal loans (Prosper, LendingClub, SocietyOne) and ride (Uber,
Lyft) to household tasks (TaskRabbit) and accommodations (Airbnb). Enabled by
the Internet and information technology, these marketplaces aim to match
sellers who are willing to share underutilized goods or services with buyers
who need them~\cite{Azevedo-matching-2016, Roth-matchmaking-2015}. Such
marketplaces are thus often referred to as examples of the so-called ``sharing
economy''~\cite{Malhotra-sharing-2014, Cusumano-sharing-2015,
Sundararajan-sharing-2016}. Airbnb, a primary example of this type of
marketplaces, connects hosts who have spare rooms with guests who need
accommodations. Founded in $2008$, it now has more than two million listings
located in more than $191$ countries, and has accumulated more than $60$
million guests.\footnote{https://www.airbnb.com/about/about-us}

Accompanied by its rapid growth, Airbnb has confronted intense debates,
regulatory challenges, and battles. Advocates argue that the platform enables
many householders to become small business owners and reduce their rental
burden. It may also make travelers pay less than hotel prices and have a more
local and authentic experience, exemplified by its slogan ``live there.''
Opponents argue that many hosts rent their entire homes for short-terms, which
is illegal in many cities~\cite{Schneiderman-Airbnb-2014, NYT-illegal-2014,
Guttentag-airbnb-2015}. This usage also involves two other critical issues.
First, entire homes that are used for short-term rentals on Airbnb may be taken
down from local housing markets, which may drive the rents up. A recent study,
however, suggested that this might not be the case~\cite{Stulberg-538-2016}.
Second, the coming-in of more travelers may be disruptive to residential
neighborhoods. Another argument from opponents is that some hosts who own a
large number of listings may be operating business on Airbnb but may fail to
fulfill their tax obligations. This gives them advantages over hotels and makes
them ``free riders,'' because accommodation taxes collected from hotels are
often used for tourism promotions that benefit all accommodation
suppliers~\cite{Guttentag-airbnb-2015}.

Although much debated, to what extent the discussed arguments are empirically
grounded remains to be seen. To this end, here we present the first large-scale
data-driven study on Airbnb, focusing on the entire market. By measuring key
characteristics directly related to these arguments, we paint a more complete
yet complicated picture of Airbnb. First, regrading the issue of short-term
rentals of entire homes, we document that across many countries, entire homes
account for the majority of listings; $68.5\%$ of all listings are entire homes
and only $29.8\%$ are private rooms. Although we do not further quantify the
extent to which they are used for short-term rentals---a limitation of our
work---simply due to the unavailability of proprietary data from Airbnb on
listing bookings, we note that the statistics of room types have changed from
$2012$ when $57\%$ are entire homes and $41\%$ are private
rooms~\cite{Guttentag-airbnb-2015}. This change suggests that Airbnb has been
becoming more like a rental marketplace rather than a spare-room sharing
platform. Moreover, listings owned by business operators may more likely be
rented in short-terms, compared with other ordinary hosts.

Second, regarding the issue of business operators, we characterize in a great
detail who they are and what their listings are. Our results suggest a heavier
usage from business operators than previously thought. The number of listings
owned by a host is distributed according to a power-law, spanning three orders
of magnitude. One third of all listings are owned by $9.4\%$ of hosts, each of
whom has at least three listings, and one host even owns $1,800$ listings.
Furthermore, we show that business operators are early-movers towards joining
Airbnb and behave more professionally than ordinary hosts and that their
listings are disproportionately of the entire home type and located in the US.
These results reinforce the rental marketplace notion of Airbnb.

Third, our analysis reveals predominantly positive star-ratings of listings,
which is different from previously observed J-shaped distribution. This
positivity bias is consistent with a bias toward using positive words in
reviews, and the extent is greater than Yelp reviews. These results may suggest
that many guests had overall positive experiences during their stays,
corroborating advocates' argument on traveler experiences. It can also indicate
the presence of selection bias in review
behaviors~\cite{Fradkin-bias-2015}---only those who had great experiences chose
to give reviews.

Taken together, we believe that our work significantly advances the current
understanding of how Airbnb is being used. Our main contributions in this work
are:
\begin{itemize}
\item We crawl Airbnb listing data on a global scale. (\S~\ref{sec:data})
\item We analyze geolocations, room types, star-ratings, and reviews of
listings. (\S~\ref{sec:listing})
\item We characterize hosts who own multiple listings on Airbnb and those
listings owned by them. (\S~\ref{sec:multi-listing})
\item We investigate factors linked to listings' future rental performance.
(\S~\ref{sec:review-growth})
\end{itemize}

\section{Data} \label{sec:data}

\subsection{Data Collection}

\begin{table}
\caption{Statistics of our data set about Airbnb}
\label{tab:stats}
\begin{tabular}{l l r}
\toprule
\multicolumn{2}{l}{Countries} & $193$ \\
\midrule
\multirow{3}{*}{Listings} & Active & $2,018,747$ \\
& Unavailable & $284,039$ \\
& Total & $2,302,786$ \\
\midrule
\multirow{3}{*}{Users} & Hosts & $1,313,626$ \\
& Guests & $11,150,017$ \\
& Total & $12,156,178^{*}$ \\
\midrule
\multicolumn{2}{l}{Reviews} & $19,377,978$ \\
\bottomrule
\textit{Note:} & \multicolumn{2}{l}{$^{*}$ $307,465$ users are both hosts and guests}
\end{tabular}
\end{table}

On Airbnb, each listing has a web page showing its details such as room type,
price, reviews from previous guests, host information, etc. For example, the
listing called ``Van Gogh's Bedroom'' can be visited at
https://www.airbnb.com/rooms/10981658. The main goal of our data collection is
to accumulate as many listings as possible, so that we can perform a systematic
analysis. There are two steps in the data collection process: (1) accumulating
listing IDs; and (2) downloading their HTML files and reviews. We describe them
in detail below, and before that, let us present in Table~\ref{tab:stats} some
summary statistics about the collected data set. To our best knowledge, these
are the first public and exact statistics about the Airbnb marketplace.

In the first step, we accumulated listing IDs by exploiting the hierarchical
structure of the Airbnb site map, which has three levels: country, regions in
the country, and search results of listings in the region. The top level at
\texttt{/sitemaps} lists $152$ countries, each of which has a hyperlink
pointing to the web page that lists regions in the country. All the regions in
Australia, for example, are listed at \texttt{/sitemaps/AU}. Each region,
again, has a hyperlink pointing to the page of search results of listings in
the region. Our script followed all these hyperlinks and obtained $83,174$
regions in $152$ countries. Then for each region, our script visited its search
page and saved all the listing IDs there. We repeated this search process $7$
times, with each new search resulting in a smaller number of new IDs. The only
constraint stopping us from more searches is the availability of computing and
storage resources. In total, we accumulated $2,302,786$ unique listings.

In the second step, for each listing, our script visited its web page, saved
the HTML file, and collected all the reviews. As we are also interested in
rental performances of listings, we repeated this step $3$ times---after the
1st, 2nd, and 7th search. The statistics in Table~\ref{tab:stats} and the
analyses presented in \S~\ref{sec:listing} and \S~\ref{sec:multi-listing} are
based on the latest crawl. We found $284,039$ ($12.3\%$) listings were
unavailable, by which we mean visiting them redirected to a search page with a
message saying ``the listing is no longer available.'' As an example, see
\texttt{/rooms/11599049}.

The entire data collection process was performed between May and September,
$2016$. Note that we followed the crawler-etiquette described in Airbnb's
\texttt{robots.txt}:\footnote{https://www.airbnb.com/robots.txt} None of the
three directories we visited---\texttt{/sitemaps}, \texttt{/s/}, and
\texttt{/rooms/\{listing ID\}}---are specified as disallowed in the file.

\subsection{Summary Statistics Analysis}

\begin{figure}
\includegraphics[width=\columnwidth]{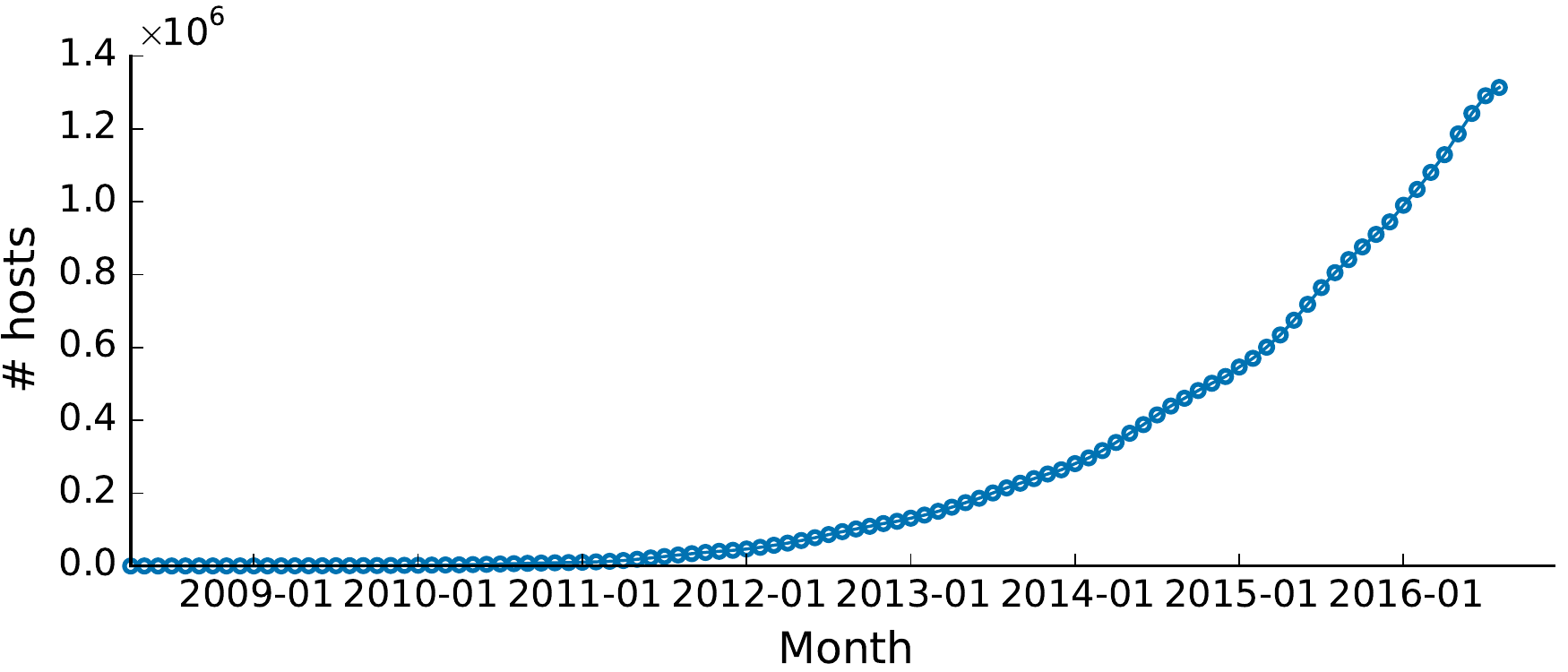}
\caption{Cumulative number of hosts.}
\label{fig:host-month}
\end{figure}

\begin{figure*}
\includegraphics[width=\textwidth]{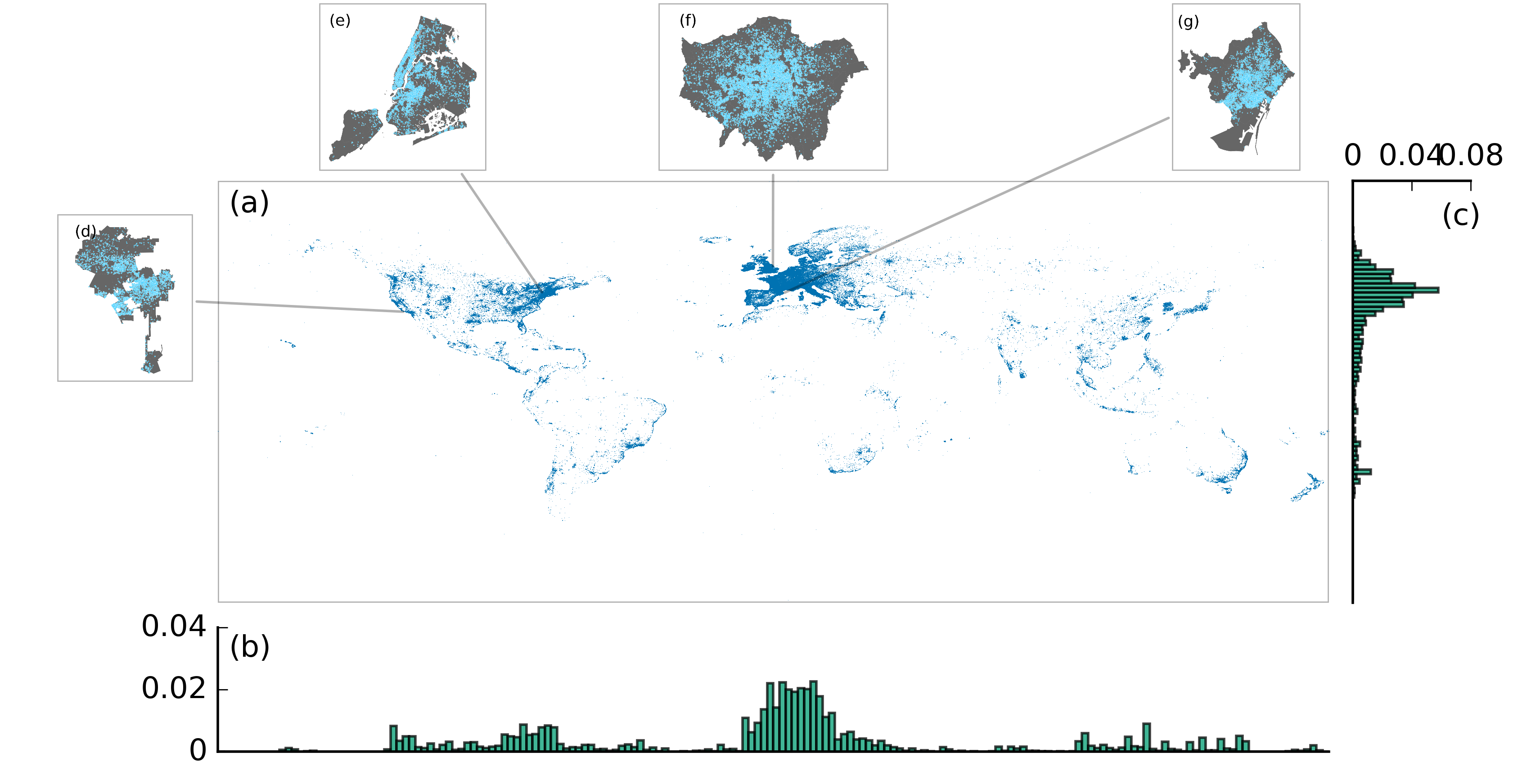}
\caption{Geolocations of Airbnb listings. (a) Dot plot of all the 2,018,747
active listings; (b--c) Histogram of longitude and latitude; (d--g) Dot plot of
listings in the city of Los Angeles, New York, London, and Barcelona. See
Appendix~\ref{app:copyright} for the copyright of city maps.}
\label{fig:listing-world}
\end{figure*}

We compare the statistics in Table~\ref{tab:stats} with official ones. The only
official but approximate numbers we found were already mentioned in
\S~\ref{sec:intro}: $191$+ countries, $2M$+ listings, and $60M$+ total guests.
Our obtained number of countries and listings seem to be comparable to official
ones. We note, however, that the number of guests ($11M$) is much smaller than
the official one ($60M$). One obvious reason is that here we have only counted
those guests who have left at least one review captured in our data set but not
every guest has given reviews.

Next, we provide estimations of some statistics that are still unknown or may
be outdated. First, $307,465$ users are both hosts and guests, accounting for
$23.4\%$ of all hosts, counted based on the observation that they own listings
and give reviews to other listings. This gives one answer to the Quora
question.\footnote{https://www.quora.com/What-percentage-of-Airbnb-hosts-are-also-guests}
Second, Fig.~\ref{fig:host-month} presents the cumulative number of hosts in
each month from March $2008$ to August $2016$, constructed using the month
information that indicates when they joined Airbnb. We see that an exponential
growth of number of hosts started from around $2012$. Third, we make an
estimate of an important statistic---the guest-to-host review rate---meaning
the fraction of stays where guests have left reviews to their hosts after the
conclusion of their stays. We approximate it as
\begin{equation}
\begin{split}
\text{guest-to-host review rate}
&= \frac{\text{\# stays with reviews left}}{\text{\# total stays}} \\
&\approx \frac{19,102,711}{102,718,148} = 18.6\%.
\end{split}
\end{equation}
Here the number of stays with reviews left is simply the number of reviews
(excluding automatic reviews, cf.~\S~\ref{subsec:review}), and the total number
of stays is estimated by extrapolating the distribution of number of reviews
per guest showed in Fig.~\ref{fig:review}(a) from the $11M$ guests to the
entire population of $60M$ guests. Our estimate has already been significantly
smaller than the reported $72\%$ in $2012$ by Airbnb's CEO Brian
Chesky.\footnote{https://www.quora.com/What-percent-of-Airbnb-hosts-leave-reviews-for-their-guests/answer/Brian-Chesky?srid=uU9cX}
It remains to be seen how accurate our estimation is and how the one disclosed
in early days is different from the one nowadays.

\section{Measuring Airbnb Listings} \label{sec:listing}

\subsection{Geolocations}

\begin{figure*}
\centering
\includegraphics[width=\textwidth]{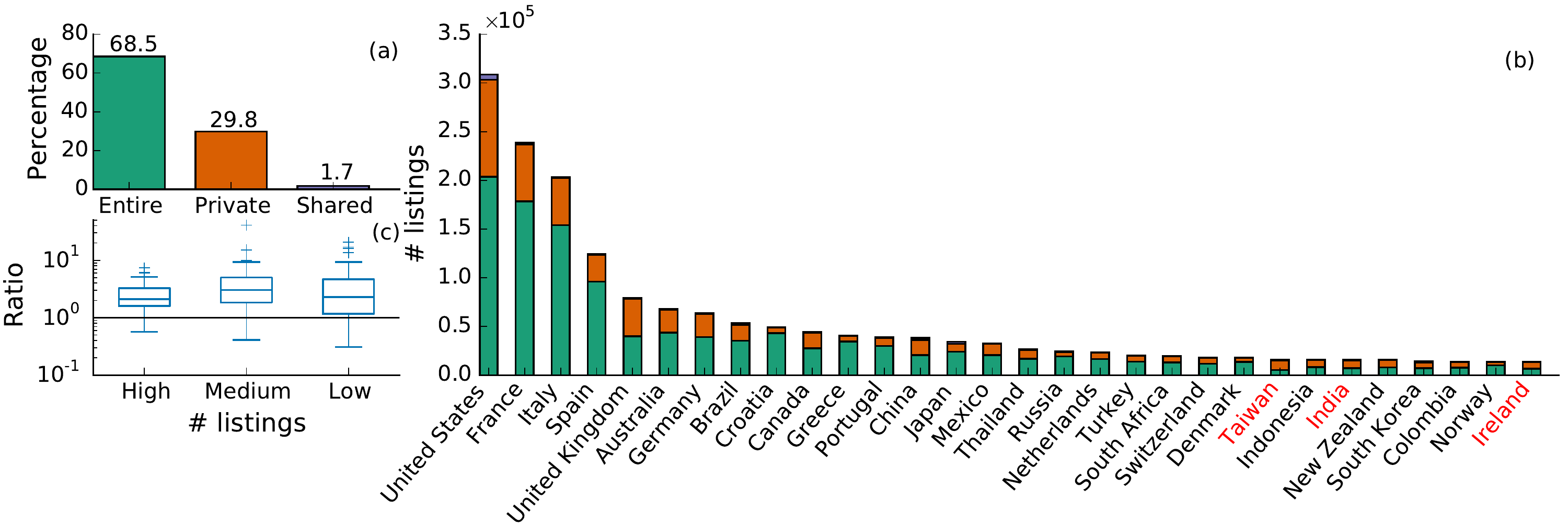}
\caption{Room types of Airbnb listings. (a) Distribution of room types of all
active listings; (b) Number of listings by room type in the top 30 countries
with the highest number of listings. Countries with smaller number of entire
homes than private rooms are marked red; (c) Distribution of the ratio between
entire homes and private rooms in countries with high, medium, and low number
of listings.}
\label{fig:rtype-country}
\end{figure*}

Let us start with where Airbnb listings are located---a question repeatedly
asked by hoteliers, policy-makers, and other stakeholders
(e.g.,~\cite{Schneiderman-Airbnb-2014}). Using each listing's approximate
geolocation information in latitude/longitude values provided by Airbnb, we
present in Fig.~\ref{fig:listing-world}(a) a dot plot showing geolocations of
all active listings across the world. We see that listings are globally
distributed. To understand their geographic concentration,
Figs.~\ref{fig:listing-world}(b) and (c) respectively show the histograms of
longitude and latitude values. We observe that, on the continental level,
listings are heavily located in Western Europe, North America, East and South
Asia, and Pacific Asia. Focusing on the country level, Airbnb has reached a
world-wide yet heterogeneous coverage. US is the largest market for Airbnb,
with $308,714$ listings totaling for $15.29\%$ of all listings, followed by
France ($11.82\%$), Italy ($10.07\%$), Spain ($6.16\%$), and United Kingdom
($3.93\%$). Figure~\ref{fig:rtype-country}(b) lists the top $30$ countries,
which in total account for $83.55\%$ of all listings. Meanwhile, many countries
have only hundreds of listings, and there are no listings in a lot of African
countries.

Focusing on cities, Figs.~\ref{fig:listing-world}(d)--(g) respectively display
locations of listings in the city of Los Angeles, New York, London, and
Barcelona. As here we are interested in the global distribution, we leave the
detailed study of how listings are located within cities as future work. Some
studies have done so focusing on London~\cite{Quattrone-who-2016} and
Barcelona~\cite{Gutierrez-BCN-2016}.

\subsection{Room Types}

Next, we study the distribution of the three types of rooms of Airbnb listings:
\textit{entire home/apartment}, \textit{private room}, and
\textit{shared room}. As suggested by their names, entire home means that the
host will not be present in the home during one's stay; private room means that
the guest will occupy a private bedroom and share other spaces with others; and
shared room means the guest will share the bedroom with other guests. While the
latter two types of listings may align with the symbolism of the sharing
economy that hosts occasionally share their spare rooms, it is the type of
entire home that (1) directly contrasts with such symbolism; and (2) becomes
the necessary condition for hosts to convert residential houses into short-term
rentals and for business operators to conduct business by renting out their
numerous properties. Hosts who use entire homes in such ways have become one of
the main targets of regulations in some cities. For instance, the New York
State Legislature recently passed a bill that subjects hosts to fines when they
rent their entire homes for less than thirty
days,\footnote{http://www.forbes.com/sites/briansolomon/2016/06/17/new-york-wants-to-fine-airbnb-hosts-up-to-7500}
and the bill was recently signed by New York Governor
Andrew~M.~Cuomo.\footnote{http://www.nytimes.com/2016/10/22/technology/new-york-passes-law-airbnb.html}
The statistics about room types are therefore among the key characteristics
mentioned in many reports by various interest groups
(e.g.,~\cite{Schneiderman-Airbnb-2014}), yet they are still unknown at the
entire-marketplace level.

Figure~\ref{fig:rtype-country}(a) shows that on the entire market, $68.5\%$ of
listings are entire homes, while only $29.8\%$ are private rooms; Airbnb has
$1.3$ times more entire homes than private rooms. These statistics are in
contrast with the ones back in $2012$ when $57\%$ were entire homes and $41\%$
were private rooms~\cite{Guttentag-airbnb-2015}. Such change indicates that
Airbnb, a primary example of the ``sharing economy,'' is more like a rental
marketplace rather than a spare-room sharing platform.

We investigate variations of the room type distribution across countries.
Figure~\ref{fig:rtype-country}(b) shows the number of the three types of
listings in the top $30$ countries with the largest number of listings. Among
them, $27$ countries have more entire homes than private rooms. In the US,
which has the largest number of listings, $65.8\%$ are entire rooms, and the
ratio between number of entire homes and private rooms reaches $2$. The only
three countries or regions where there are more private rooms than entire homes
are Taiwan ($0.57$), India ($0.91$), and Ireland ($0.96$), though the ratio is
close to one for the latter two countries. We further calculate the ratio for
each of the $150$ countries with more than $100$ listings (Countries with small
number of listings have large fluctuations of the ratio.), and show in
Fig.~\ref{fig:rtype-country}(c) the distributions of the ratio for the three
equally-sized groups of countries, based on their total number of listings. We
see that the ratio is greater than one across many countries and even larger
for countries with smaller number of listings.

\subsection{Star-ratings}

We now focus on star-ratings and reviews, which are the reputation system of
Airbnb and important sources of information for guests to pick
listings~\cite{Luca-design-2016}. At the conclusion of each stay, both the host
and guest can give reviews to and rate each other at a scale from $1$ to $5$
stars with a unit of $0.5$ star. Each listing will receive an average
star-rating once it is rated by at least $3$
guests.\footnote{https://www.airbnb.com/help/article/1257/how-do-star-ratings-work}
The star-rating of each individual review, however, is not publicly disclosed.

\begin{figure}
\centering
\includegraphics[width=\columnwidth]{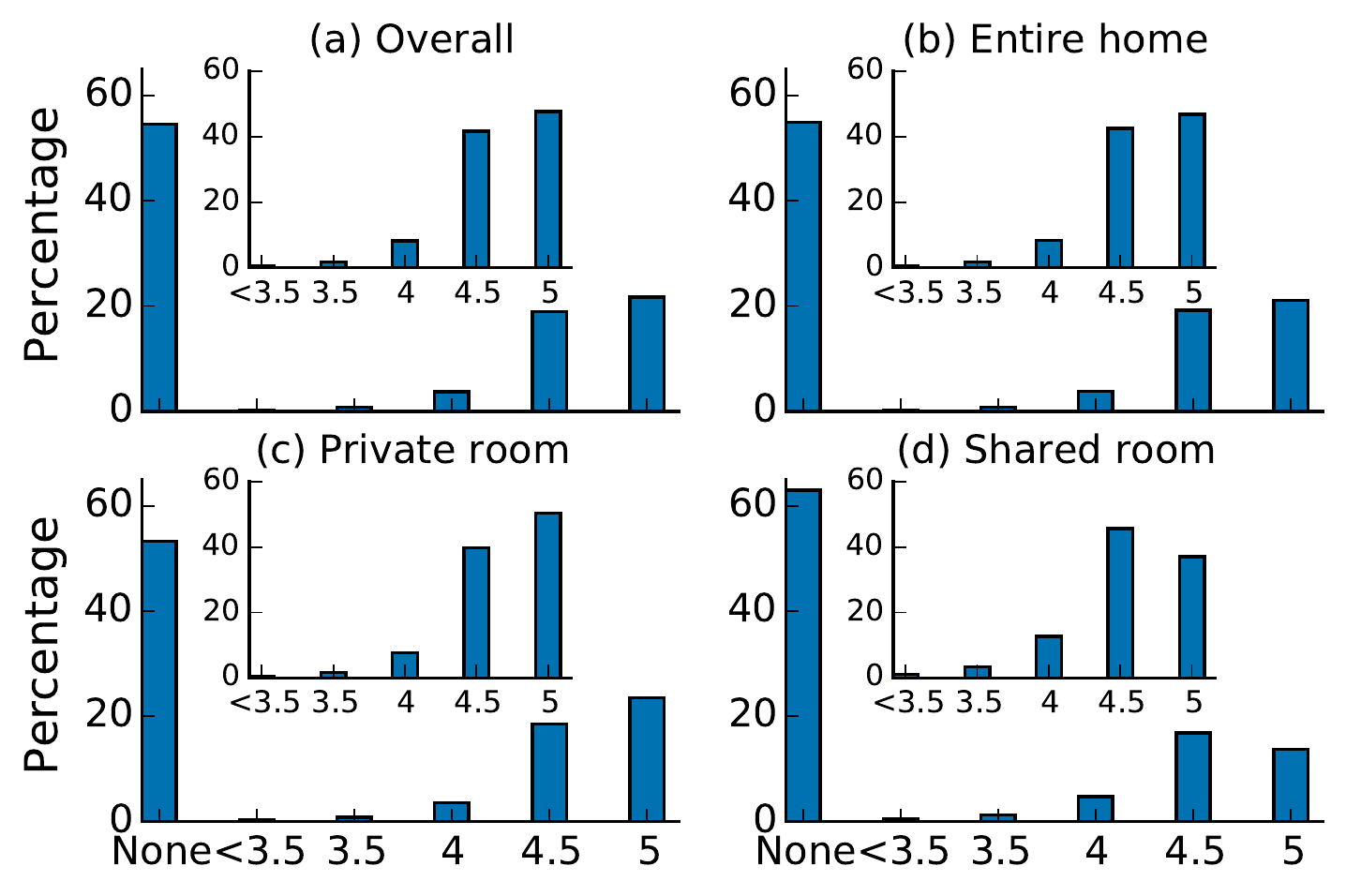}
\caption{Distribution of star-ratings.}
\label{fig:rating-rtype}
\end{figure}

Figure~\ref{fig:rating-rtype}(a) shows a bimodal distribution of star-ratings
over all listings. More than half of them ($54.6\%$) have not received their
ratings, and $40.6\%$ have $4.5$ or $5$ stars. These three categories of
listings account for over $95\%$ of all listings, and the number of listings
with $3.5$ or lower stars is essentially negligible. Focusing on listings that
have received star-ratings, Fig.~\ref{fig:rating-rtype}(a) inset shows that
star-ratings are overwhelmingly positive; $89.5\%$ of them have $4.5$ or $5$
stars, and the mean (median) rating is $4.67$ ($4.5$). These results are
consistent with a previous small-scale study~\cite{Zervas-review-2015}.
However, this heavily skewed distribution is sightly different from previously
observed J-shaped distribution of product reviews~\cite{Hu-Jshape-2009}. In
particular, such distribution suggests that the number of 1-star products is
high, which is not the case for Airbnb listings.

We explore variations of this distribution by room type.
Figures~\ref{fig:rating-rtype}(b)--(d) respectively show the star-rating
distributions of entire homes, private rooms, and shared rooms. Although, for
shared rooms, $4.5$-star listings are of a higher fraction than $5$-star
listings, there is very limited variation of the distribution: The majority of
listings, irrespective of room type, have either no or very high ratings. This,
on the one hand, suggests that many guests had great experiences during their
stays, but makes star-ratings less informative and distinguishable for
future guests to choose among potential listings, on the other.

\subsection{Reviews} \label{subsec:review}

There are $19,377,978$ reviews given by $11,150,017$ guests. We are aware that
Airbnb will post an automatic review if a host cancels a reservation, serving
as one penalty for the
cancellation.\footnote{https://www.airbnb.com/help/article/314/why-did-i-get-a-review-that-says-i-canceled}
We find $275,267$ automatic reviews, amounting to $1.4\%$ of all reviews. This
provides an upper bound of the cancellation rate by hosts, as not every stay
yields a review. We removed all automatic reviews before further analysis.

\subsubsection{Distribution of Review Counts}

\begin{figure}
\includegraphics[width=\columnwidth]{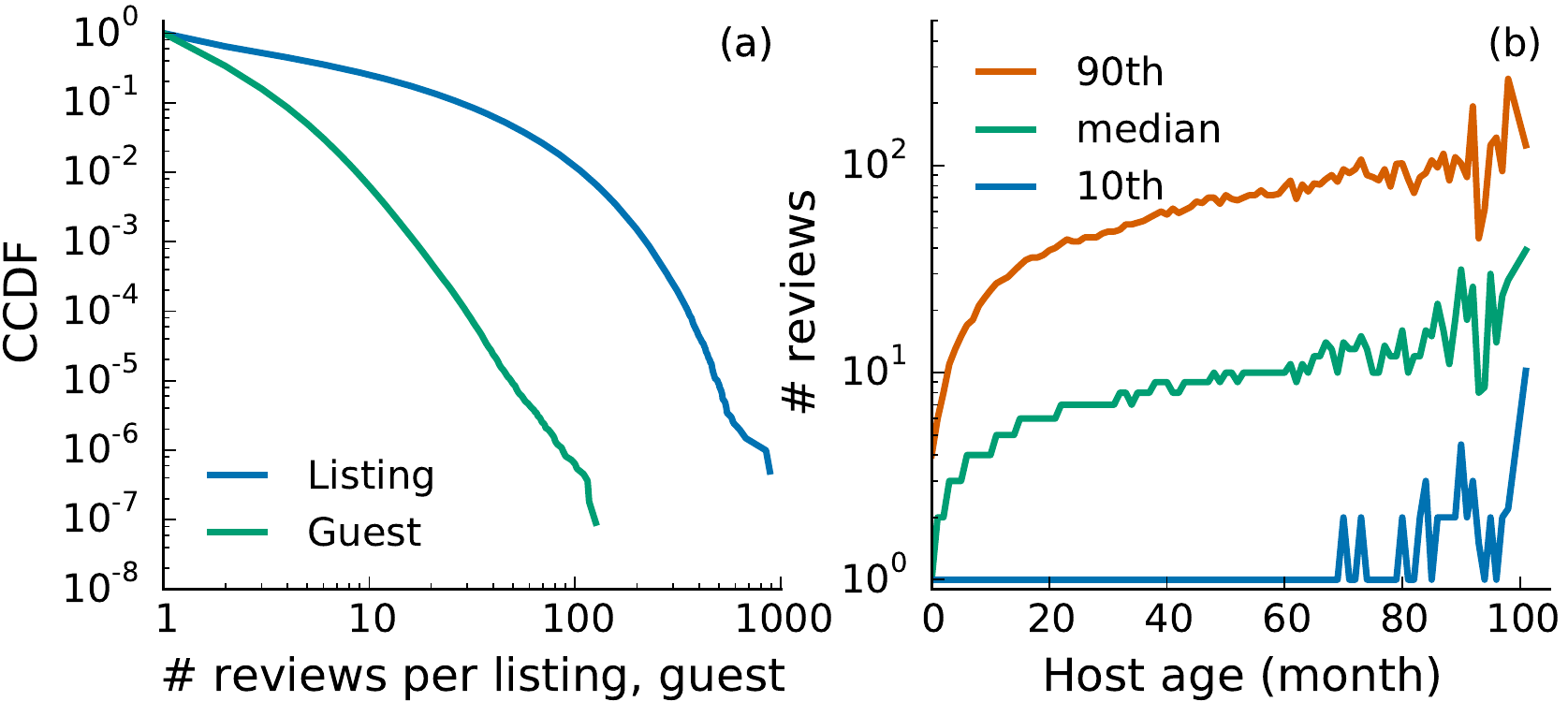}
\caption{Review counts on Airbnb. (a) Distribution of number of reviews per
listing and guest; (b) Number of reviews of listings with different host ages.}
\label{fig:review}
\end{figure}

On the Airbnb review system, which is different from others like Yelp, only
guests who concluded their stays can give reviews. This makes the number of
reviews a listing has received a proxy of its business attention, although the
review rate and number of stayed nights associated with each review may be
different. Therefore, we analyze how reviews are distributed among listings,
showing in Fig.~\ref{fig:review}(a) the survival distribution. We shift it by
one to make the zero-review data point visible in the logarithmic scale. We see
that although Airbnb is a relatively young marketplace, reviews have already
been heterogeneously distributed among listings. About $35.7\%$ listings do not
have reviews, and respectively $12.2\%$ and $7.4\%$ listings have one and two
reviews. The remaining $44.6\%$ listings, each of which has at least three
reviews, account for $93.6\%$ reviews. In \S~\ref{sec:review-growth}, we
demonstrate the presence of the rich-get-richer mechanism in explaining the
growth of reviews.

We next analyze the relation between listing age and number of reviews. As we
do not know when each listing was established, nor do we know when each and
every review was given, we use the Airbnb age of its host---the number of
months passed since they joined Airbnb---as a proxy of the listing age. Focusing
on listings with at least one review, Fig.~\ref{fig:review}(b) shows the 10th,
50th, and 90th percentile of number of reviews of listings grouped by host age.
We observe that (1) the number of reviews in general increases with host ages;
(2) even for hosts who joined Airbnb for years, the median number of reviews
still remains in the order of ten; and (3) the review count is heterogeneously
distributed even for hosts who joined Airbnb in the same month.

Figure~\ref{fig:review}(a) also shows a heavy-tailed distribution of number of
reviews per guest. Respectively $66.3\%$ and $17.8\%$ guests have given one and
two reviews, while only $0.63\%$ guests have left at least $10$ reviews.

\subsubsection{Review Content} \label{subsubsec:word}

We start investigating text content of reviews with what languages are used.
Using the \texttt{langdetect} language detection
library,\footnote{https://pypi.python.org/pypi/langdetect} we found $49$
languages used. Table~\ref{tab:lang} reports the percentages of reviews written
in the top $10$ most used languages. English dominates this ranking, with
$72.8\%$ of reviews using it, followed by French, Spanish, German, and Italian.
From now on, we focus on the $14,094,229$ English reviews.

\begin{table}
\caption{Top 10 languages used in reviews}
\label{tab:lang}
\begin{tabular}{l r | l r}
\toprule
Language & $\%$ & Language   & $\%$ \\
\midrule
English  & 72.8 & Chinese (Simplified) & 1.6 \\
French   & 10.3 & Korean     & 1.3 \\
Spanish  & 3.8  & Portuguese & 1.0 \\
German   & 3.5  & Dutch      & 0.9 \\
Italian  & 2.3  & Russian    & 0.9 \\
\bottomrule
\end{tabular}
\end{table}

\textbf{Positive/Negative words:} Recall that a vast majority ($89.5\%$) of
listings have $4.5$ or $5$ stars (Fig.~\ref{fig:rating-rtype}(a) inset). This
raises the question of whether this strongly skew toward positive star-ratings
is consistent with a usage bias toward positive vocabulary. To answer this, we
use a recently released resource that contains norms of almost $14K$ English
words~\cite{Warriner-norms-2013}, each of which has a valence score from $1$ to
$9$, where valence greater than $5$ means positive words and smaller than $5$
negative words. We calculate the ratio of the frequency of positive and
negative words in Airbnb reviews. To compare the skewness, we also calculate
the ratio for $2.7M$ Yelp
reviews.\footnote{https://www.yelp.com/dataset\_challenge/}
Table~\ref{tab:ratio} shows that the ratio doubles for Airbnb reviews. This
confirms a bias toward using positive words, and the extent is even greater
than reviews on Yelp, which already exhibits a positive
bias~\cite{Jurafsky-yelp-2014}. Given that Airbnb is a P2P platform where hosts
can also choose which guests to accommodate, this finding may open up further
investigations into user behaviors on different platforms.

\section{Measuring Multi-listing Hosts} \label{sec:multi-listing}

In this section, we investigate a key issue repeatedly discussed in the current
debate---the existence of hosts who own multiple listings on Airbnb. Hereafter,
we call them ``multi-listers.'' Note that they may have different names in
various reports, such as ``commercial hosts,'' ``professional hosts,'' and
``business operators,'' all attempting to capture the possibility that they may
operate business on Airbnb, as an ordinary host is less likely to own numerous
listings. Despite being a critical issue, there has been no systematic analysis
about multi-listers and their listings.

\subsection{Existence of Multi-listers}

Recall that there are $2,018,747$ active listings owned by $1,313,626$ hosts
(Table~\ref{tab:stats}), thus on average every host owns $1.54$ listings.
Figure~\ref{fig:nlisting-th}(a) shows the survival distribution of number of
owned listings per host. We observe a somewhat surprising heavy-tailed
distribution that spans more than three orders of magnitude, similar to what
have been observed in many complex systems~\cite{Albert-cn-2002}. We fit the
empirical distribution with a power-law function
$p(x) = \frac{\alpha-1}{x_{\min}} \left( \frac{x}{x_{\min}} \right)^{-\alpha}$
using the methods developed in refs~\cite{Clauset-powerlaw-2009,
Alstott-powerlaw-2014}, and obtain $\alpha=2.65$ and $x_{\min}=15$. These
results not only demonstrate the existence of ``super multi-listers''---hosts
who can own up to $1,800$ listings, but also indicate that the existence of
multi-listers is prevalent. Simply put, although the vast majority of hosts
have a small number of listings, there is a consistent number of hosts who own
a large number of listings. In particular, $1,030,134$ ($78.4\%$) and $159,627$
($12.2\%$) hosts respectively own one and two listings, and the remaining
$9.43\%$ hosts own $33.16\%$ of all listings.

\begin{table}
\caption{Ratio of frequency of positive and negative words used in Airbnb and
Yelp reviews}
\label{tab:ratio}
\begin{tabular}{l c c}
\toprule
      & Airbnb reviews & Yelp reviews \\
\midrule
Ratio & 13.749         & 6.705 \\
\bottomrule
\end{tabular}
\end{table}

To further characterize multi-listers and their listings, we need to answer a
key question---which threshold of number of owned listings per host allows us
to separate hosts into two groups and then compare them. Previous literature
have not reached a consensus about this. The New York State Attorney General,
for example, defined ``commercial hosts'' as those who have three or more unique
listings~\cite{Schneiderman-Airbnb-2014}. Li \textit{et al.} defined
``professional hosts'' as those who have two or more
listings~\cite{Li-agent-2015}. Here, we argue that a typical threshold value
may not be well-defined, as Fig.~\ref{fig:nlisting-th}(a) clearly suggests that
the number of owned listings is a multi-scale phenomenon. Moreover, as we shall
show, focusing on a particular value loses the whole picture and can even be
misleading.

Instead, we simply increase the threshold and study how various measures of
interest change accordingly. Specifically, for a given threshold, we
characterize (1) the subset of hosts whose number of owned listings
\textit{exceeds} the given threshold; and (2) the subset of listings owned by
those hosts. If the threshold is zero, we simply focus on all hosts and all
listings.

\subsection{Listings Owned by Multi-listers}

\begin{figure*}
\includegraphics[width=\textwidth]{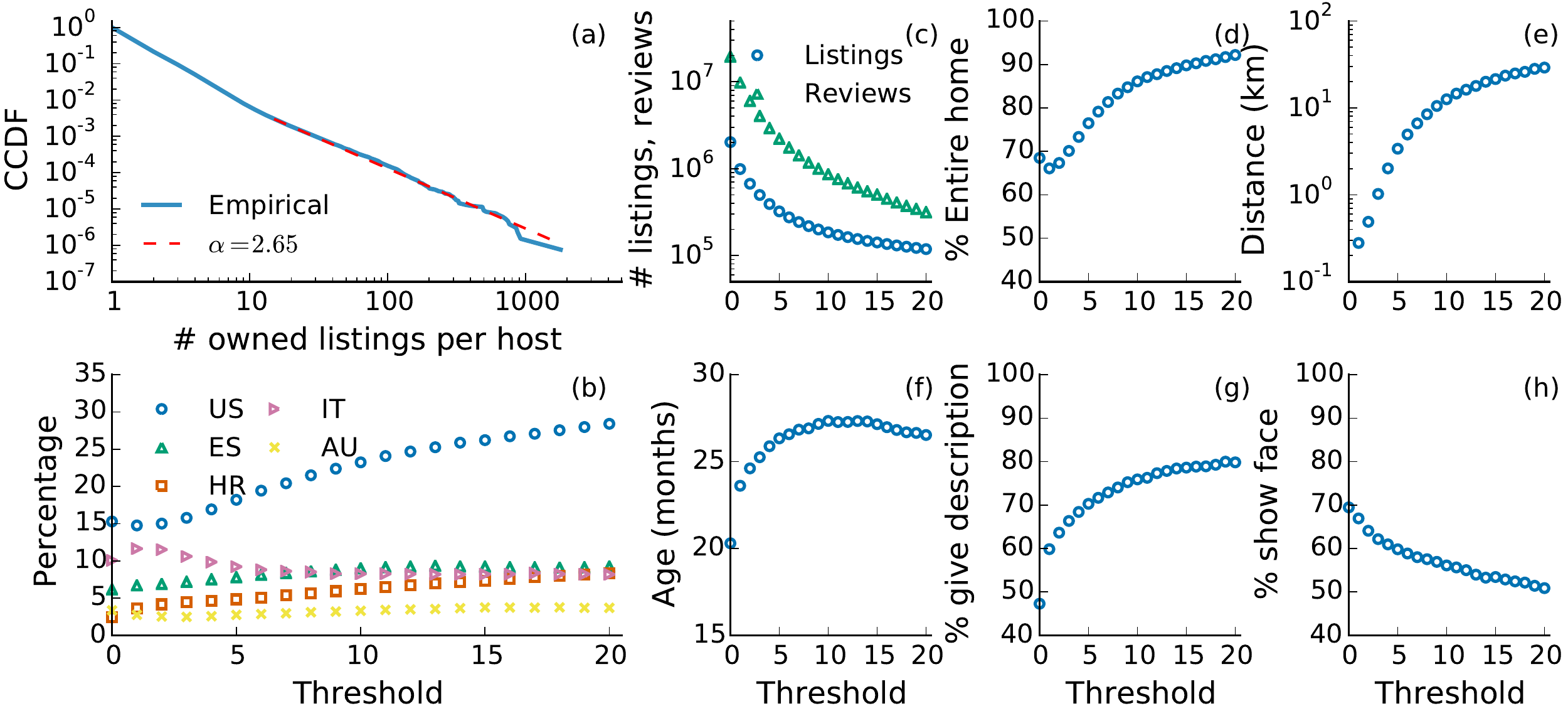}
\caption{Characterizing multi-listers and their listings. (a) Survival
distribution of number of owned listings per host. From (b) to (h), we focus on
(1) the subset of hosts whose number of owned listings \textit{exceeds} a
threshold; and (2) the subset of listings owned by those hosts, and show how
various measures change as we increase the threshold. (b) Percentage of
listings located in 5 countries; (c) Number of listings and their total
reviews; (d) Percentage of entire homes; (e) The median of the distribution of
the maximum distance between pairs of listings owned by a host; (f) Mean Airbnb
age of hosts; (g) Percentage of hosts who give descriptions; (h) Percentage of
hosts whose profile photos have facial features.}
\label{fig:nlisting-th}
\end{figure*}

Figures~\ref{fig:nlisting-th}(b)--(e) present the characterization results for
listings owned by multi-listers. Figure~\ref{fig:nlisting-th}(b) shows the
percentages of listings in $5$ countries, United States (US), Spain (ES),
Croatia (HR), Italy (IT), and Australia (AU), selected because they have the
largest number of listings when the threshold is $20$. We observe that as we
increase the threshold, listings owned by multi-listers are disproportionately
located in the US. We also see that a decreasing portion of listings are from
Italy, while Spain and Croatia have an increasing portion.
Figure~\ref{fig:nlisting-th}(b) (and Fig.~\ref{fig:nlisting-th}(d)) also
illustrates why focusing on a particular threshold can be misleading. For
example, if one focused on a particular value ($1$ or $2$), one would have
concluded that the number of US listings owned by multi-listers is proportional
to total listings in the US, which is not the case.

Figure~\ref{fig:nlisting-th}(c) focuses on how the total number of listings and
number of reviews received by them change as we increase the threshold. We
observe a faster decrease of review counts, indicating that the average number
of reviews per listing decreases. When the threshold is $0$, on average a
listing has $9.6$ reviews, which decreases to $2.6$ when the threshold reaches
$20$. Therefore, listings owned by multi-listers are less reviewed.

Figure~\ref{fig:nlisting-th}(d) shows that an increasing portion of listings
are entire homes as the threshold increases. When the threshold is $20$, more
than $92\%$ of listings are entire homes, compared to $68.5\%$ on the entire
market. This confirms the previous conjecture that commercial hosts seek to
rent out their entire home properties.

Figure~\ref{fig:nlisting-th}(e) answers the question of how far away the
listings owned by a single host. Using listings' latitude/longitude values, we
calculate, for each host, the maximum distance among all pairwise distances
between two of their listings, capturing the geographical diameter of their
``managerial'' activities. Figure~\ref{fig:nlisting-th}(e) shows that the median
of the distribution of maximum distance over all hosts whose number of listings
exceeds a given threshold, though keeps increasing, is in the order of $10km$.
This suggests that listings owned by multi-listers may locate within a city and
that they may operate business locally.

\subsection{Multi-listers}

Figures~\ref{fig:nlisting-th}(f)--(h) show the characterization results for
multi-listers. First, Fig.~\ref{fig:nlisting-th}(f) demonstrates that
multi-listers are early-movers towards joining Airbnb, as their mean Airbnb age
is larger than other hosts.

Figure~\ref{fig:nlisting-th}(g) focuses on the percentages of hosts who give
descriptions in the space-limited ``Your Host'' section on listings' web pages.
Presenting self-description is one important way to establish trust between
guests and hosts. Surprisingly, only $47.3\%$ of all hosts have given
descriptions (threshold $0$), and multi-listers are more likely to do so. Using
the method described in ref.~\cite{Monroe-fighting-2008} and setting the
threshold to $10$, we find the top $10$ overrepresented words used in
multi-listers' descriptions are ``https,'' ``vacation,'' ``wildsch\"{o}nau,''
``properties,'' ``rentals,'' ``team,'' ``villas,'' ``rental,'' ``apartments,'' and
``services,'' indicating that they use the description section to advertise
their listings.

Figure~\ref{fig:nlisting-th}(h) examines another way to establish
trust---showing faces in profile photos. Using a service provided by Face++ to
detect whether there are facial features presented in a given
photo,\footnote{http://www.faceplusplus.com/detection\_detect/} we find that
for $69.5\%$ of all hosts, there are faces detected in their profile photos.
Multi-listers are less likely to show faces in their photos, and a manual
inspection reveals that many of them use company's logos as profile photos.

\section{Modeling Review Growth} \label{sec:review-growth}

In the last section, we have found notable differences between multi-listers
and other hosts. This raises the question of whether the differences are linked
to listings' future rental performances. To answer this, we approximate
performances with number of new reviews, as we do not know listings' actual
booked nights. For each listing, we calculate the number of new reviews it has
received in one month, which is our response variable. The first two columns in
Table~\ref{tab:regress} list predictors and their definitions. Instant Book is
a listing feature, meaning that a potential guest can book the listing without
the host's approval.\footnote{https://www.airbnb.com/help/article/187/what-is-instant-book}
The \texttt{superhost} badge is awarded to a host if they satisfies a series of
requirements set by Airbnb.\footnote{https://www.airbnb.com/superhost} Response
time of a host is transformed into numerical values, so that a faster response
corresponds to a larger value.

We fit a linear regression model by ordinary least squares (OLS) to understand
factors linked to the growth of reviews. The last column in
Table~\ref{tab:regress} presents the regression results. Other predictors being
the same, listings with more existing reviews will have more new reviews,
demonstrating the presence of the rich-get-richer mechanism that has been found
to explain the growth of numerous systems~\cite{Perc-Matthew-2014}. Listings
whose ratings have one more star will have $0.169$ more review. A private room
will gain $0.112$ more review than an entire home, while a shared room will
have $0.091$ less review than an entire home. A listing that can be instantly
booked will have $0.272$ more review than those without the Instant Book
feature. Being a \texttt{superhost}, giving descriptions, maintaining high
response rate, responding in short time, and owning a smaller number of
listings all have positive effects on review growth, though the effect is small
for the number of owned listings.

\begin{table}
\caption{Regression results for monthly new reviews}
\label{tab:regress}
\begin{tabular}{l l l}
\toprule
$reviews$ & Number of reviews & $0.037^{***}$ \\
& & (0.0001) \\
$rating$ & Star-rating & $0.169^{***}$ \\
& & (0.001) \\
$room\_type$ & Room type & \\
& \hspace{2cm}private & $0.112^{***}$ \\
& & (0.003) \\
& \hspace{2cm}shared  & $-0.091^{***}$ \\
& & (0.011) \\
$amenities$ & \# available amenities & $0.017^{***}$ \\
& & (0.0003) \\
$instant\_book$ & Instant book is allowed & \\
& \hspace{2cm}1 & $0.272^{***}$ \\
& & (0.003) \\
$photos$ & Number of photos & $-0.002^{***}$ \\
& & (0.0001) \\
$host\_age$ & Host age & $-0.018^{***}$ \\
& & (0.0001) \\
$host\_super$ & Host is a \texttt{superhost} & \\
& \hspace{2cm}1 & $0.259^{***}$ \\
& & (0.005) \\
$host\_desp$ & Host gives descriptions & \\
& \hspace{2cm}1 & $0.040^{***}$ \\
& & (0.003) \\
$host\_resp\_rate$ & Host response rate & $0.229^{***}$ \\
& & (0.013) \\
$host\_resp\_time$ & Host response time & $0.220^{***}$ \\
& & (0.002) \\
$host\_nlisting$ & \# owned listings & $-0.001^{***}$ \\
& & (0.00002) \\
Constant & & $-0.446^{***}$ \\
 & & (0.011) \\
\midrule
Observations & \multicolumn{2}{r}{1,140,488} \\
Adjusted R$^{2}$ & \multicolumn{2}{r}{0.388} \\
\bottomrule
\textit{Note:} & \multicolumn{2}{r}{$^{*}$p$<$0.1; $^{**}$p$<$0.05; $^{***}$p$<$0.01} \\
\end{tabular}
\end{table}

\section{Related Work}

There is a growing interest in Airbnb and other sharing economy platforms from
diverse disciplines, ranging from computer science to economics to law.
Empirical studies have focused on, for example, star-ratings of
listings~\cite{Zervas-review-2015} and geolocations of listings within
cities~\cite{Quattrone-who-2016, Gutierrez-BCN-2016}. Our focus here is the
entire Airbnb platform. There are studies that have reported the presence of
discrimination on Airbnb~\cite{Edelman-digital-2014, Edelman-racial-2017}. Some
work have investigated factors associated with listings' price, such as the
receipt of star-ratings~\cite{Gut-price-2015} and race~\cite{Kakar-race-2016}
and personal photos~\cite{Ert-trust-2016} of their hosts. The impact of Airbnb
on hotel industry revenue~\cite{Zervas-impact-2015, Zervas-impact-2016} and on
tourism industry employment~\cite{Fang-effect-2016} has been investigated.
Discussions about regulations of Airbnb have also generated much
attention~\cite{Cohen-self-2015, Edelman-shortcuts-2015}. Li \textit{et~al.}
investigated differences in performances and behaviors between professional
hosts---those who own two or more listings---and non-professional hosts on
Airbnb~\cite{Li-agent-2015}. Our analysis, however, reveals the lack of
threshold that allows us to separate professional and non-professional hosts.
Studies focusing on the motivations behind joining Airbnb to provide
hospitality have pointed out that monetary compensation and sociability are two
important aspects~\cite{Ikkala-monetizing-2015, Lampinen-hosting-2016}. Fradkin
\textit{et~al.} experimentally investigated the determinants and bias in the
Airbnb review system~\cite{Fradkin-bias-2015}. Fradkin proposed ranking
algorithms for the Airbnb search engine~\cite{Fradkin-search-2015}.

\section{Conclusion and Future Work}

In this work, we have presented the first large-scale data-driven study on
Airbnb. After crawling the largest ever number of Airbnb listings, we have
measured their geolocations, room types, star-ratings, and reviews. We have
also characterized in a great detail hosts who own multiple listings as well as
their listings. We have built a linear regression model to understand factors
linked to the growth of reviews. As these aspects are among the key points
discussed in the ongoing debate and among important features in the sharing
economy, we believe that our work provides valuable insights for various
stakeholders and may serve as a public and empirical reference to inform the
debate.

One major limitation of our work is that we do not measure listing occupancy.
Therefore we do not know to what extent a listing is rented in short-terms or
what revenue differences are between multi-listers and ordinary hosts. We
notice that it is feasible to crawl listing calendar data. However, one issue
of using such data is that, given a listing is unavailable on some dates, we
cannot tell if it is rented out or simply blocked by the host for not renting.
Another technical challenge is to large-scale monitor the calender data on a
daily basis. Future work may do so on a small scale. Other future work include
studying geolocation at a finer level, measuring hosts' behavioral changes, and
understanding the effect of review content on listings' future rentals.

\appendix

\section{Copyright of City Maps} \label{app:copyright}

\textbf{Los Angeles}: shapefile from \url{https://goo.gl/SxFhCx}.
\textbf{New York City}: shapefile from \url{https://goo.gl/Isb4rf}.
\textbf{London}: shapefile from \url{https://goo.gl/bWn9ua}; copyright:
``Contains National Statistics data \textcopyright Crown copyright and database
right [2015]'' and ``Contains Ordnance Survey data \textcopyright Crown
copyright and database right [2015]''. \textbf{Barcelona}: shapefile from
\url{https://goo.gl/bYOhWL}.

\begin{acks}
I thank the anonymous referees for their comments and suggestions and
\href{http://cnets.indiana.edu/}{Center for Complex Networks and Systems
Research} and \href{http://www.soic.indiana.edu/}{School of Informatics and
Computing} at Indiana University for excellent computing resources.
\end{acks}

\end{document}